\documentclass{article}
\usepackage{spconf,amsmath,graphicx}
\usepackage{hyperref, adjustbox, booktabs, amsfonts, aligned-overset, stmaryrd}

\newcommand{\tinyrightarrow}{\scriptscriptstyle\shortrightarrow}
\newcommand{\tinyleftarrow}{\scriptscriptstyle\shortleftarrow}
\newcommand{\overbar}[1]{\mkern 1.5mu\overline{\mkern-1.5mu#1\mkern-1.5mu}\mkern 1.5mu}

\title{Dual-path Mamba: Short and Long-term Bidirectional Selective Structured State Space Models for Speech Separation}
\name{Xilin Jiang, Cong Han, Nima Mesgarani}
\address{Department of Electrical Engineering, Columbia University, USA}
\begin{document}
\ninept
\maketitle
\begin{abstract}
Transformers have been the most successful architecture for various speech modeling tasks, including speech separation. However, the self-attention mechanism in transformers with quadratic complexity is inefficient in computation and memory. Recent models incorporate new layers and modules along with transformers for better performance but also introduce extra model complexity. In this work, we replace transformers with Mamba, a selective state space model, for speech separation. We propose dual-path Mamba, which models short-term and long-term forward and backward dependency of speech signals using selective state spaces. Our experimental results on the WSJ0-2mix data show that our dual-path Mamba models of comparably smaller sizes outperform state-of-the-art RNN model DPRNN, CNN model WaveSplit, and transformer model Sepformer. Code: https://github.com/xi-j/Mamba-TasNet
\end{abstract}
\section{Introduction}
\label{sec:intro}

Speech is an inherently long signal of thousands of samples. Speech separation models separate multiple individual speech, each of thousands of samples, from a single overlapped speech mixture of the same length. Therefore, an effective mechanism for modeling long speech sequences is crucial to achieving high performance in speech separation. With the advancement of deep learning, a variety of neural sequence modeling architectures, including convolutional neural networks (CNN) \cite{cnn}, recurrent neural networks (RNN) \cite{rnn}, and transformers \cite{transformer}, have been adopted into speech separation models and achieved a state-of-the-art (SOTA) performance at the time they were adopted \cite{convtasnet, dprnn, sepformer}. However, each architecture faces its own challenges: CNNs are restricted by a finite receptive field; RNNs are hard to parallelize and struggle with vanishing or exploding gradients \cite{rnngradients}; Transformers often outperform CNNs and RNNs but require quadratic computation in computing self-attention \cite{transformer}. Although alternative self-attention mechanisms with sub-quadratic complexity exist, they fail to match the performance of the original attention mechanism in speech separation \cite{explore_sepformer}. Meanwhile, the attempt to scale up transformer parameters, as seen from Sepformer \cite{sepformer} to almost eight times larger QDPN \cite{QDPN}, yields a modest performance improvement, suggesting diminishing returns with increasing model size. Recently, MossFormer2, which integrates transformers and RNN-free recurrent networks, has achieved better performance but is much smaller than QDPN, hinting that speech separation might still benefit from exploring beyond transformer architectures.

State space model (SSM) \cite{kalman, gu2021combining, s4} is another unique class of sequence modeling architecture. Recently, a novel selective SSM named Mamba is proposed \cite{mamba}. Distinct from earlier SSMs, Mamba incorporates an input-dependent selection mechanism that improves sequence modeling performance but still enjoys linear complexity with respect to the sequence length. Mamba models have matched and even surpassed transformers of comparable sizes in sequence modeling tasks of text, audio, image, and genomics \cite{mamba, vision_mamba}.

Inspired by Mamba's effectiveness and efficiency in sequence modeling, we introduce it into speech separation and propose a new model named \textit{dual-path Mamba} (DPMamba). As the name suggests, we follow the long sequence modeling method in dual-path RNN \cite{dprnn} to split a long speech into multiple short chunks and apply Mamba models within each chunk, across all chunks, in the original direction, and in the reversed direction of time. Our experiments of DPMamba of three different sizes (XS, S, M) on the WSJ0-2mix dataset \cite{wsj2mix} demonstrate on-par or superior performance over SOTA models of similar or larger sizes, including the CNN-based Wavesplit \cite{wavesplit}, the RNN-based DPRNN \cite{dprnn} and VSUNOS \cite{vsunos}, and the transformer-based DPTNet \cite{dptnet} and SepFormer \cite{sepformer}. We are scaling up the model parameters and will report the performance of our largest model (L) soon.

\begin{figure*}[!ht]
  \centering  \includegraphics[width=0.975\textwidth]{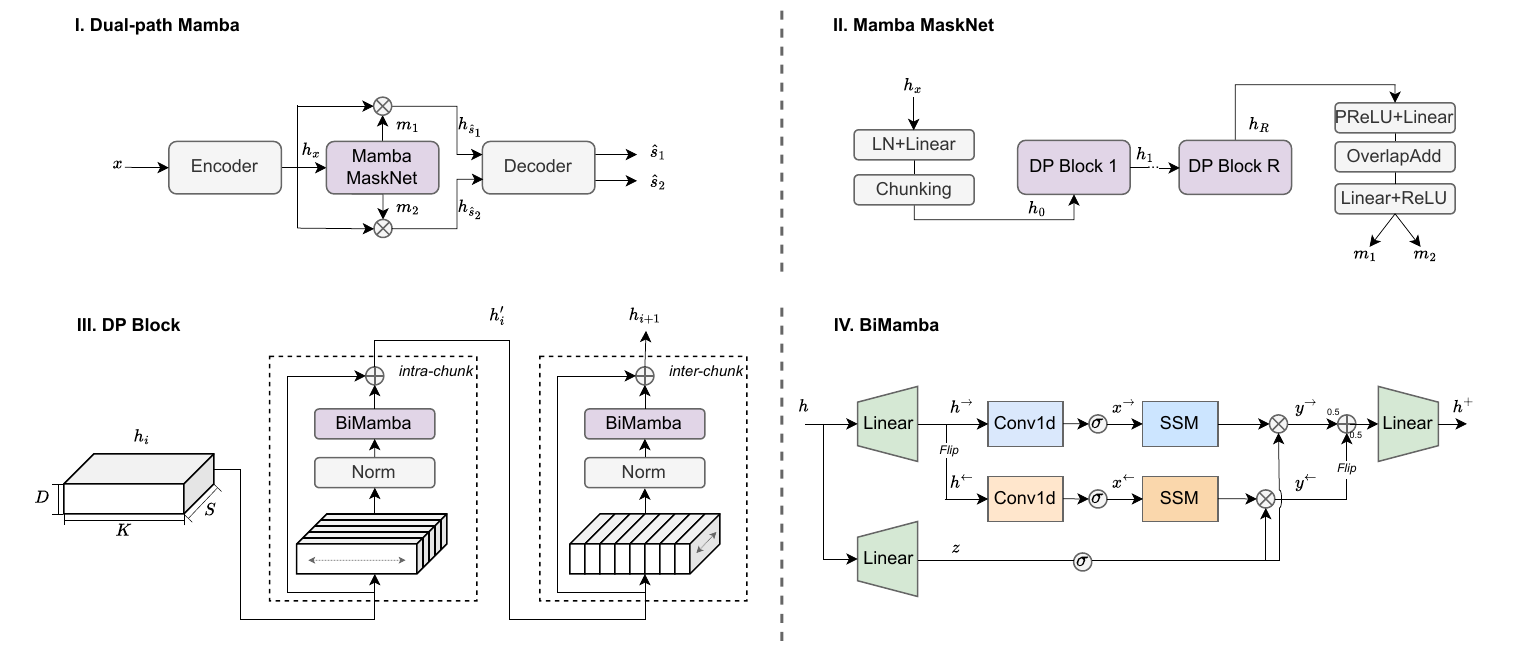}
  \caption{A top-down view of DPMamba from I to IV.}
  \label{fig:dpmamba}
\end{figure*}

\section{Related Works}

\subsection{Speech Separation} \label{sec:related_models}

Following the shift to adaptive front-ends over the short-time Fourier transform for waveform encoding \cite{adaptive, tasnet}, time-domain speech separation models have become the mainstream of research. The core of these models is a mask estimation network which consists of multiple separation blocks of the same architecture. The architecture of these blocks plays a crucial role in separation performance and has thus become a primary focus of research efforts. CNN and RNN models were first introduced, succeeded by better transformer and hybrid models. We spotlight models that have set benchmarks in speech separation across different architecture families and will compare them with our mamba separation models in Section \ref{results}. 

\noindent\textbf{CNN}: Conv-TasNet \cite{convtasnet} first outperforms ideal time-frequency masking in speech separation. It uses temporal convolutional networks (TCN) with progressively increasing dilation factors to expand its receptive field. Sudo rm -rf \cite{sudormrf, sudormrf2} employs a more efficient CNN architecture with down-sampling layers to expand the receptive field. Wavesplit \cite{wavesplit} builds on TCNs but introduces additional speaker embeddings, which leads to a better performance.

\noindent\textbf{RNN}: An earliest time-domain model TasNet \cite{tasnet} utilizes long short-term memory (LSTM) to separate speech. To improve the modeling ability of long sequence, DPRNN \cite{dprnn} proposes splitting a speech sequence into smaller chunks and utilizing two LSTMs for intra-chunk and inter-chunk processing. VSUNOS \cite{vsunos} outperforms DPRNN by substituting LSTMs with gated RNNs.

\noindent
\textbf{Transformer}: DPTNet \cite{dptnet} and Sepformer \cite{sepformer} adopt the dual-path architecture of DPRNN and replace LSTMs with transformers. They differ in the stride size and the use of a RNN or positional encoding to embed the order information of a speech sequence. QDPN \cite{QDPN} proposes an improved quasi-dual-path architecture and scales up parameters to 200M. This marks the best performance of the largest transformer model in WSJ0-2mix. 

\noindent
\textbf{Hybrid}: More recent separation models employ a hybrid architecture which often augments a vanilla transformer using another model, like convolution-augmented transformers in TD-Conformer \cite{td_conformer}, gated transformer with convolution-augmented attentions in Mossformer \cite{mossformer}, and transformer and RNN-free recurrent network in Mossformer2 \cite{mossformer2}. Separate and Diffuse \cite{separate_and_diffuse} adds a generative model to post-refine the speeches separated by Sepformer.

\subsection{Applications of Mamba}
Mamba has been proven a transformer-level performance in multiple modalities and tasks. Mamba is first applied to text, audio, and genomics modeling tasks in its own paper \cite{mamba}. Later applications extend to images \cite{vision_mamba}, biomedical data \cite{ma2024u}, graphs \cite{wang2024graph}, motions \cite{zhang2024motion}, videos \cite{yang2024vivim}, and point clouds \cite{liang2024pointmamba}. For speech, \cite{quan2024multichannel} adopts Mamba for multi-channel speech enhancement. \cite{s4m} adopts S4 (the predecessor of Mamba) \cite{s4} for single-channel speech separation. To our knowledge, ours is the first work that adopts Mamba in single-channel speech separation. 

\section{DPMamba}

In this section, we will first review the selective state space model which is the core component of Mamba. Then, we will explain DPMamba in a top-down way with Fig. \ref{fig:dpmamba}. We focus on our adoption of Mamba (purple parts in Fig. \ref{fig:dpmamba}) into separation since the rest is commonly shared with other dual-path separation models.

\subsection{Selective State Space Model} \label{sec:ssm}

A state space model (SSM) performs a sequence-to-sequence mapping $\textbf{x}(t) \in \mathbb{R} \mapsto \textbf{y} (t) \in \mathbb{R}$ with a hidden state $\textbf{h} \in \mathbb{R}^{H}$, a state transition matrix $\textbf{A} \in \mathbb{R}^{H \times H}$, an input projection matrix $\textbf{B} \in \mathbb{R}^{H \times 1} $, and an output projection matrix $\textbf{C} \in \mathbb{R}^{1 \times H}$. $H$ is the dimension of the hidden state:
\begin{align}
    \textbf{h}^{\prime}(t) = \textbf{A}\textbf{h}(t) + \textbf{B} \textbf{x}(t), \hspace{4pt} \textbf{y}(t) = \textbf{C}\textbf{h}(t) \label{eq:cont}
\end{align}
To compute Eq.\ref{eq:cont} for discrete-time signals in digital devices, we need to discretize the SSM with discretized matrices $\overbar{\textbf{A}}$ and $\overbar{\textbf{B}}$:
\begin{align}
    \textbf{h}_t = \overbar{\textbf{A}}\textbf{h}_{t-1} + \overbar{\textbf{B}} \textbf{x}_t, \hspace{4pt} \textbf{y}_t = \textbf{C}\textbf{h}_t \label{eq:discrete}
\end{align}
$\overbar{\textbf{A}}$ and $\overbar{\textbf{B}}$ are approximated by zero-order hold. A learnable parameter \textbf{$\Delta$} balances how much to focus or ignore the current state and input:
\begin{align}
    \overbar{\textbf{A}} = \text{exp}(\textbf{$\Delta$}\textbf{A}), \hspace{4pt} \overbar{\textbf{B}} = (\textbf{$\Delta$}\textbf{A})^{-1}(\text{exp}(\textbf{$\Delta$}\textbf{A}) - I) \cdot \textbf{$\Delta$} \textbf{B}
\end{align}
By unrolling the computation in Eq.\ref{eq:discrete} along the sequence, it can be observed that the output sequence $\textbf{y}_t$ is the input sequence $\textbf{x}_t$ convolved with a structured kernel $\overbar{\textbf{K}}$ made up by $\overbar{\textbf{A}}$, $\overbar{\textbf{B}}$, and $\textbf{C}$:
\begin{align}
    \overbar{\textbf{K}} = (C\overbar{\textbf{B}}, \overbar{\textbf{A}\textbf{B}}, ..., C\overbar{\textbf{A}}^{L-1}\overbar{\textbf{B}}), \hspace{4pt} \textbf{y} = \textbf{x} \ast \overbar{\textbf{K}} \label{eq:kernel}
\end{align}
The kernel $\overbar{\textbf{K}}$ can be pre-computed if $\overbar{\textbf{A}}$, $\overbar{\textbf{B}}$, and $\textbf{C}$ are fixed. However, the SSM in Mamba is input \textit{selective} to enable input content-awareness: It updates the parameters $\textbf{$\Delta$}_t$ (and therefore $\overbar{\textbf{A}}_t$), $\overbar{\textbf{B}}_t$, and $\textbf{C}_t$ based on the input $\textbf{x}_t$ at each timestep $t$. To accommodate these dynamic updates, \cite{mamba} proposes an efficient selective scan algorithm  for the calculation of Eq.\ref{eq:kernel}.

\subsection{Time-domain Dual-path Model}
DPMamba follows the time-domain dual-path structure of previous SOTA separation models \cite{dprnn, sepformer, mossformer}. A linear encoder encodes a single-channel wave mixture $x \in {\mathbb{R}^{1 \times T}}$ of $T$ samples to a two-dimensional latent representation $h_x \in {\mathbb{R}^{D \times N}}$ of $D$ dimensions and $N$ frames. The encoder has a kernel size of $16$ and a stride of $8$, resulting in $N$= $T/8$. The encoder dimension $D$ is a hyperparameter. Next, for two speakers in the mixture, a Mamba masking network (Mamba MaskNet) estimates two masks $m_1, m_2 \in {\mathbb{R}^{D \times N}}$. These masks are then applied through element-wise multiplication with $h_x$ to isolate individual speech sources $h_{\hat{s}_1}, h_{\hat{s}_2} \in {\mathbb{R}^{D \times N}}$. Finally, a linear decoder decodes $h_{\hat{s}_1}, h_{\hat{s}_2}$ back to waveforms $\hat{s}_1, \hat{s}_2 \in {\mathbb{R}^{1 \times T}}$. Fig. \ref{fig:dpmamba} I draws the same workflow.

Mamba MaskNet is a dual-path network that operates on a three-dimensional tensor $h_0 \in {\mathbb{R}^{D \times K \times S}}$. $h_0$ is obtained by \textit{Chunking} $N$ frames of $h_x$ into multiple smaller chunks. $K$=250 is the chunk size, and $S$ is the resulting number of chunks, with an overlap factor of $50\%$ between adjacent chunks. Mamba MaskNet processes within- and across-chunk sequences of much smaller lengths $K$ and $S$ than $N$, which are two substantially easier sequence modeling tasks \cite{dprnn}. The output of Mamba MaskNet, $h_R \in {\mathbb{R}^{D \times K \times S}}$, is converted back to two-dimensional by \textit{OverlapAdd} operation. Mask $m_1$ and $m_2$ are estimated by a linear layer followed. Additional linear layers are placed before \textit{Chunking} and \textit{OverlapAdd} operation, as drawn in Fig. \ref{fig:dpmamba} II. 

Note that we use the same chunk size, kernel size, and stride as other dual-path models like Sepformer \cite{sepformer} and Mossformer \cite{mossformer, mossformer2} for a fair comparison.

\subsection{Dual-path Bidirectional Mamba}
Mamba MaskNet comprises a stack of $R$ dual-path (DP) blocks iteratively processing the chunked features $h_0$, resulting in $h_1$ to $h_R$ of the same shape. Within each DP block, four SSMs process the features in four different ways: intra-chunk forward,  intra-chunk backward, inter-chunk forward, and inter-chunk backward. An \textit{intra-chunk} SSM locally processes all $K$ frames within a chunk; An \textit{inter-chunk} SSM globally processes all $S$ chunks of the entire signal. A \textit{forward} SSM processes in the original direction; A \textit{backward} SSM processes in the opposite direction of the sequence. As shown in Fig. \ref{fig:dpmamba} III and IV, a DP block contains an intra-chunk and an inter-chunk unit interleaved, and an intra-chunk or inter-chunk unit each contains a forward and a backward SSM running in parallel.

\subsubsection{Local and Global}

We adopt computation units of the same structure for local and global processing as shown in Fig. \ref{fig:dpmamba} III. Both intra-chunk and inter-chunk units contain a normalization layer, a bidirectional Mamba (BiMamba), and a skip connection. By default, we use RMSNorm \cite{rmsnorm} as the normalization method due to its computational efficiency. An intermediate feature $h_i \in \mathbb{R}^{D \times K \times S}$ is first processed by the intra-chunk unit across $K$ frames in every chunk and then processed by the inter-chunk unit across $S$ chunks in total. The computation in $i$th DP block can be expressed as follows:
\begin{align}
    h_i^{\prime} &= h_i + \overset{\longleftarrow K \longrightarrow}{\textit{BiMamba}}\bigl(
    \textit{Norm}(h_i)\bigr) \label{eq:intra} \\
    h_{i+1} &= h_i^{\prime} + \overset{\longleftarrow S \longrightarrow}{\textit{BiMamba}}\bigl(\textit{Norm}(h_i^{\prime})\bigr) \label{eq:inter}
\end{align}

The intra-chunk duration is around 30 ms with our default configuration for 8 kHz speech. In implementation, we permute the last two dimensions $K$ and $S$ of $h_i$ ($h_i^{\prime}$) to alter between intra-chunk or inter-chunk processing.

\subsubsection{Forward and Backward}

The original Mamba \cite{mamba} uses one SSM to process the input sequence in the forward direction. However, it has been shown that bidirectional models, which utilize future context, usually outperform unidirectional models in speech separation \cite{convtasnet}. To make Mamba bidirectional, we borrow the BiMamba design from \cite{vision_mamba} to run one forward and one backward SSM in parallel, as shown in Fig. \ref{fig:dpmamba} IV. For clarity of discussion, we omit the subscripts used in the previous sections and denote the input and output of BiMamba as $h$ and $h^{+}$, respectively.

We start with an input sequence $h \in \mathbb{R}^{D \times L}$, where the dimension $D$ is the same as the encoder dimension and the length $L \in \{K, S\}$ is either the number of frames per chunk or the number of chunks. A linear layer projects $h$ to $h^{\tinyrightarrow} \in \mathbb{R}^{E \times L}$, where $E = 2D$ is the dimension expanded by 2. $h^{\tinyleftarrow}$ is $h^{\tinyrightarrow}$ with $L$ samples flipped. Meanwhile, another linear layer projects $h$ to $z \in \mathbb{R}^{E \times L}$ in order to later gate SSM outputs:
\begin{align}
    h^{\tinyrightarrow} &= \textit{Linear}_i(h), \hspace{3pt} h^{\tinyleftarrow} = \textit{Flip}(h^{\tinyrightarrow}) \\
    z &= \textit{Linear}_g(h)
\end{align}
The forward and the backward sequence $h^{\tinyrightarrow}$ and $h^{\tinyleftarrow}$ are then processed by their own convolution followed by activation layers, resulting in $\textbf{x}^{\tinyrightarrow}$ and $\textbf{x}^{\tinyleftarrow}$. The convolution has a kernel size of 4, and the activation $\sigma$ is the Sigmoid Linear Unit (SiLU) \cite{silu1, silu2}:
\begin{align}
    \left\{ 
    \begin{aligned}
    \hspace{5pt} \textbf{x}^{\tinyrightarrow} &= \sigma\bigl(\textit{Conv1d}_{\tinyrightarrow}(h^{\tinyrightarrow})\bigr) \\
    \hspace{5pt} \textbf{x}^{\tinyleftarrow} &= \sigma\bigl(\textit{Conv1d}_{\tinyleftarrow}({h}^{\tinyleftarrow})\bigr)
    \end{aligned}
    \right.
\end{align}
$\textbf{x}^{\tinyrightarrow}$ and $\textbf{x}^{\tinyleftarrow}$ are the inputs to two SSMs detailed in Section \ref{sec:ssm}. The SSM outputs are gated by $\sigma(z)$:
\begin{align}
    \left\{ 
    \begin{aligned}
    \hspace{5pt} \textbf{y}^{\tinyrightarrow} &= \sigma(z) \otimes \textit{SSM}_{\tinyrightarrow}(\textbf{x}^{\tinyrightarrow}) \\
    \hspace{5pt} \textbf{y}^{\tinyleftarrow} &= \sigma(z) \otimes \textit{SSM}_{\tinyleftarrow}({\textbf{x}}^{\tinyleftarrow})
    \end{aligned}
    \right.
\end{align}
Finally, the final output $h^{+}$ is obtained by a linear projection of the average of the forward and the backward processed sequences $\textbf{y}^{\tinyrightarrow}$ and $\textbf{y}^{\tinyleftarrow}$, after the latter flipped back to the original direction:
\begin{align}
    h^{+} = \textit{Linear}_o\bigl(\frac{\textbf{y}^{\tinyrightarrow} + \textit{Flip}(\textbf{y}^{\tinyleftarrow})}{2}\bigr)
\end{align}

\section{Results} \label{results}

\subsection{Experiments}

We implemented DPMamba of three different sizes as documented in Table \ref{table:config}. We trained them with scale-invariant signal-to-noise ratio (SI-SNR) \cite{sisnr} and evaluated them with the improvement of SI-SNR (SI-SNRi) and the improvement of signal-to-distortion ratio (SDRi) on the WSJ0-2mix dataset \cite{wsj2mix}. We trained all models with a batch size of 1, an Adam optimizer \cite{adam}, a peak learning rate of $1.5e^{-4}$, a linear learning rate warmup of 20,000 steps (1 epoch), a cosine learning rate decay to $0.01$ from the peak, and a total of 200 epochs. We apply speed perturbation with a random ratio between 95\% and 100\% and dynamic mixing as data augmentation. Note that this training setting is similar to the one in Sepformer \cite{sepformer} but with a different learning rate schedule. All the experiments are conducted in a NVIDIA L40 GPU with mixed (bfloat16) precision.

\begin{table}[!t]
    \caption{Hyperparameters of DPMamba XS, S, M, and L. $R \times 2$ means $R$ DP blocks and each block consists of one intra-chunk and one inter-chunk BiMamba unit.\\}
    \begin{adjustbox}{width=0.8\columnwidth,center}
    \begin{tabular}{c|cccc}
    \toprule
    Model  & Dimension $D$  & \#Layers & \#Params (M) \\ \hline \hline
    DPMamba (XS) & 128 & 8 $\times$ 2  & 2.3 \\
    DPMamba (S) & 256 & 8 $\times$ 2   & 8.1 \\
    DPMamba (M) & 256 & 16 $\times$ 2  & 15.9 \\
    \bottomrule
    \end{tabular}
    \end{adjustbox}
    \label{table:config}
    \vspace{-10pt}
\end{table}

\begin{table}[t]
    \caption{A comparison of DPMamba with previous SOTA separation models from each architecture family on WSJ0-2mix. n.r. stands for \textit{not reported}.\\}
    \begin{adjustbox}{width=\columnwidth,center}
    \begin{tabular}{c|cccc}
    \toprule
    Model  & SI-SNRi (dB) & SDRi (dB) & \#Params (M) & Stride \\ \hline \hline
    \textit{CNN} & & & & \\
    Conv-TasNet \cite{convtasnet} & 15.3 & 15.6  & 5.1  & 8   \\
    Sudo rm -rf (B=36) \cite{sudormrf2} & 19.5 & n.r.  & 23.2  & 10   \\
    Wavesplit \cite{wavesplit} & 22.2 & 22.3  & 29  & 1   \\ \hline
    \textit{RNN} & & & & \\
    TasNet \cite{tasnet} & 10.8 & 11.1  & n.r.  & 20   \\
    DPRNN \cite{dprnn} & 18.8 & 19.0  & 2.6  & 1   \\
    VSUNOS \cite{vsunos} & 20.1 & 20.4  & 7.5  & 2   \\ \hline
    \textit{Transformer} & & & & \\
    DPTNet \cite{dptnet} & 20.2 & 20.6  & 2.6  & 1   \\
    Sepformer \cite{sepformer} & 22.3 & 22.4  & 25.7  & 8   \\
QDPN \cite{QDPN} & 23.6 & n.r.  & 200  & 8   \\ \hline
    \textit{Hybrid} & & & & \\
    Mossformer (L) \cite{mossformer} & 22.8 & n.r.  & 42.1  & 8   \\
    TD-Conformer-XL \cite{td_conformer} & 21.2 & n.r.  & 102.7  & 8  \\
    Separate And Diffuse \cite{separate_and_diffuse} & 23.9 & n.r.  & n.r.  & 8  \\
    Mossformer2 (L) \cite{mossformer2} & 24.1 &  n.r. & 55.7  & 8   \\ \hline
    \textit{SSM} & & & & \\
    S4M-tiny \cite{s4m} & 19.4 & 19.7 & 1.8  & 8   \\
    S4M \cite{s4m} & 20.5 & 20.7 & 3.6  & 8   \\
    DPMamba (XS) & 19.2 & 19.4 & 2.3  & 8   \\
    DPMamba (S) & 21.4 & 21.6 & 8.1  & 8  \\
    DPMamba (M) & 22.6 & 22.7 & 15.9  & 8   \\ 
    \hline
    \end{tabular}
    \end{adjustbox}
    \label{table:all}
     \vspace{-10pt}
\end{table}

\subsection{Performance Comparison}

In Table \ref{table:all}, we compare the performance of our models with the performance of the SOTA models from different architecture families: \textit{CNN}, \textit{RNN}, \textit{Transformer}, and \textit{Hybrid}, which we have introduced in Section \ref{sec:related_models}. We also include the performance of S4M \cite{s4m}, which adopts the S4 model for separation and also belongs to the SSM family. The numbers in the table correspond to their best models trained with data augmentation, as reported in their papers.

Our three DPMamba models, spanning from XS of 2.3 million parameters to M of 15.9 million parameters, achieve SI-SNRi values between 19.2 dB and 22.6 dB and SDRi values from 19.4 dB to 22.7 dB \footnote{There was an error in our training and evaluation code. We have fixed it and updated the results from the previous version. We are training and will report the performance of our largest model DPMamba (L) soon.}. Notice that DPMamba models achieve a high performance with a relatively smaller model size or larger stride, compared to models from other families. For instance, DPMamba (XS), despite having slightly fewer parameters and a stride of 8, outperforms DPRNN, which uses a stride of 1. A larger stride means a larger down-sampling ratio by the encoder, leading to faster inference and smaller memory usage. DPMamba (S) is marginally larger than VSUNOS but achieves an SI-SNRi 1.3 dB higher, with our stride 4 times larger. DPMamba (M) outperforms Sepformer and is only 0.2 dB lower than Mossformer in SI-SNRi, however, it only requires around 60\% and 40\% of their parameters, respectively. Finally, our S and M models outperform the previous SSM model S4M, although the XS model slightly lags behind the performance of S4M-tiny. The S4M paper \cite{s4m} does not implement models of comparable sizes to our S and M, preventing a conclusive comparison across all model sizes. 

\begin{figure}[!ht]
  \centering  \includegraphics[width=0.8\linewidth]{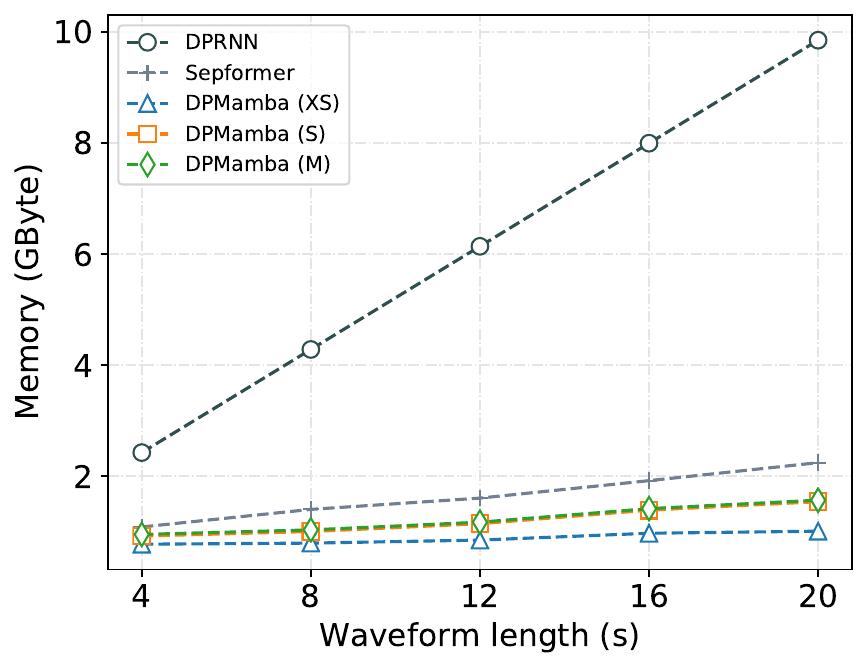}
  \caption{A comparsion of GPU memory usage of DPMamba with Sepformer and DPRNN.}
  \label{fig:memory}
\end{figure}

In Fig. \ref{fig:memory}, we benchmark the GPU memory consumption during a forward pass of the models on waveforms of different lengths. We compare our models against Sepformer and DPRNN with default model configurations implemented by Speechbrain \cite{speechbrain}. DPMamba (M) maintains a Sepformer-level performance but reduces memory usage by 30\%. DPMamba (XS) outperforms DPRNN and consumes only 10\% of memory in separating 10-second speeches. The small memory usage marks the potential of our model in mobile devices.

\subsection{Ablations}

We conducted ablation studies of the DPMamba (S) models. These involved switching from bidirectional to unidirectional models, decreasing or increasing the dimension $H$ of the hidden state of SSMs, eliminating dynamic mixing in training, and replacing RMSNorm with LayerNorm. We trained all model for 100 epochs. 

Our numbers in Table  \ref{table:ablation} reveal that a variation in the hidden state dimension above 8 and the choice between RMSNorm and LayerNorm have little impact on the separation performance. Training with dynamic mixing boosts both the SI-SNRi and SDRi by 0.6 dB. The most significant performance improvement, a boost of over 3 dB, comes from adding backward SSMs in the bidirectional model, compared to only using forward SSMs in the unidirectional model.

\begin{table}[t]
    \caption{Ablations on DPMamba (S).}
    \begin{adjustbox}{width=0.9\columnwidth,center}
    \begin{tabular}{c|ccc}
    \toprule
    Configuration  & SI-SNRi (dB) & SDRi (dB) & \#Params (M) \\ \hline \hline
    %Single-path & . & .  & 7.9  \\
    Unidirectional & 16.9 & 17.2  & 7.4  \\
    LayerNorm & 20.6 & 20.8  & 8.1  \\ \hline
    H=8 & 20.6 & 20.8  & 7.7  \\
    H=32 & 20.6 & 20.8  & 8.9  \\ \hline
    Without DM & 20.0 & 20.2  & 8.1  \\ \hline
    Default (S, 100 epochs) & 20.6 & 20.8  & 8.1 \\ \hline
    \end{tabular}
    \end{adjustbox}
    \label{table:ablation}
\end{table}

\section{Conclusion}

In this work, we introduce DPMamba, a new model for speech separation. DPMamba utilizes a dual-path network to model local and global aspects of speech sequences and incorporates bidirectional Mamba blocks for processing the sequences in forward and backward directions. Our models of three different sizes either meet or surpass the performance of existing CNN, RNN, and transformer models of similar or large sizes. Moving forward, we will explore two directions: enhancing the efficiency of the Mamba separation model and improving the performance by integrating Mamba with other network layers.

\section{Acknowledgement}

This work was funded by the National Institutes of Health (NIH-NIDCD) and a grant from Marie-Josee and Henry R. Kravis. 

\bibliographystyle{IEEEbib}
\bibliography{strings,refs}

\end{document}